\documentclass[twocolumn,aps,pra]{revtex4}
\usepackage{amsmath}
\usepackage{graphicx}
\usepackage{algorithm}
\usepackage{algpseudocode}
\usepackage{physics}

\begin{document}

%Title of paper
\title{Hyperspectral Reconstruction using Discrete LED-Structured Illumination}

\author{John C. Howell}
\affiliation{Institute for Quantum Studies, Chapman University, Orange, CA 92866, USA}
\affiliation{Racah Institute of Physics, The Hebrew University of Jerusalem, Jerusalem, Israel, 91904}

\author{Pieter H. Neethling}
\affiliation{Stellenbosch Photonics Institute, Stellenbosch University, Stellenbosch, South Africa}
\affiliation{National Institute for Theoretical and Computational Sciences, South Africa}

\author{Tjaart P. J. Krüger}
\affiliation{Department of Physics, University of Pretoria, Pretoria, South Africa}
\affiliation{National Institute for Theoretical and Computational Sciences, South Africa}
\affiliation{Forestry and Agricultural Biotechnology Institute (FABI), University of Pretoria, Pretoria, South Africa}

\email{johhowell@chapman.edu}

\begin{abstract}
We consider the use of digital signal processing to reconstruct continuous reflectance spectra using a small finite set of randomly illuminated light emitting diodes (LEDs).  We simulate the use of LEDs having identical spectral distance and Gaussian bandwidth whose illumination overlaps its nearest neighbors.  An object, whose reflectance spectrum is to be determined, is illuminated by a series of random spectral patterns consisting of randomly chosen LEDs with random intensity. We quantify the information within the illumination patterns using the singular value decomposition (SVD) and reconstruct reflectance spectra, specifically hemoglobin and several green vegetation spectra using the pseudoinverse of the SVD for a given amount of noise. We show that for sparse plant spectra, it is possible to reconstruct the continuous green vegetation spectra with RMSE less than 1\% with as few as 25 LEDs. Our study demonstrates that reconstructing sparse reflectance spectra based on random structured illumination can enable low-cost LED-based cameras to perform equally well as expensive cameras, especially for dedicated applications.
            
\end{abstract}

\maketitle
\section{Introduction}
Multispectral and hyperspectral imaging are used in a wide array of technological domains \cite{calvini2019growing} including but not limited to precision agriculture \cite{avola2023overview}, medicine \cite{fei2019hyperspectral}, geological and mineralogical exploration \cite{peyghambari2021hyperspectral}, environmental monitoring \cite{podorozhniak2021usage}, detection of defects in meats \cite{feng2018hyperspectral}, pollution detection \cite{kar2016classification}, forensic analysis \cite{pallocci2022forensic,de2023hyperspectral}, and quality control.  Multispectral imaging typically captures between 2 and 15 different colors per image pixel while hyperspectral imaging can capture hundreds or even thousands of contiguous colors per pixel. Each imaging modality has its advantages and disadvantages that are usually in direct conflict \cite{han2017overview}. For example, the reduced number of color channels for multispectral imaging requires less data storage, but often lacks sufficient detail for  classification or material analysis and vice versa for hyperspectral images.  

Due to the high costs of commercial cameras, which makes them inaccessible to many research laboratories and small companies, several research groups have explored the use of LEDs for multispectral reflectance imaging owing to their low cost, light weight, small size, low power consumption, flexibility, programmability, and wide range of available wavelengths \cite{li2012multi, shrestha2013multispectral, Herrera-Ramirez2014, goel2015hypercam, salvador2018low, modir2022led}. The authors believe this is an important domain of research, having recently developed a low-cost, LED-based, Raspberry Pi-driven multispectral camera as an educational and research tool for developing countries \cite{howell2024raspberrypimultispectralimaging}. Since LEDs emit relatively narrow spectral bands, continuous broadband reconstruction requires the use of several LEDs, usually with spectral interpolation, which enhances the technical complexity and therefore requires adequate digital signal processing. 

Structured spectral illumination is an attractive method for general spectral reconstruction because it enables the recovery of spectral shapes with fewer measurements, higher robustness, and flexible system design. LED-based structured illumination is frequently used for colorimetric reconstruction of images, whereby sequential or random time-multiplexed illumination is used \cite{Park2007, shrestha2013multispectral}. The latter illumination strategy involves switching random subsets of LEDs on simultaneously. Examples include computational ghost-imaging variants with LED arrays and several compressive hyperspectral microscopy works~\cite{Huang2022}. However, for spectral reconstruction, only sequential illumination has thus far been used to the best of our knowledge~\cite{Herrera-Ramirez2014, goel2015hypercam, Tschannerl, Parmar}, 
often with low precision and/or computationally expensive algorithms \cite{Xiao2019, Ma2022tongue, Zhang2021}. 

The use of random structured spectral illumination allows the use of the mathematics of random-measurement matrices for reconstruction of sparse spectra. These random matrices can be used in popular digital signal processing techniques like compressive sampling \cite{donoho2006compressed,baraniuk2007compressive,candes2008introduction,eldars2012compressed}, which also rely on sparsity and measurement incoherence.  

In this study, we use random time-multiplexed spectral illumination for the reconstruction of reflectance spectra, which enabled us to use compressive sensing reconstruction models. We also vary the total number of LEDs to determine the optimal number and bandwidths of LEDs for reconstructing given reflection spectra. We study random structured spectral illumination using a small set of relatively broadband LEDs for sparse reflectance spectra.  The implementation of this illumination scheme should be relatively straightforward, as illumination setups with similar capabilities have previously been used for compressive spectral imaging \cite{Wang2023} and reflectance reconstruction using a programmable LED illuminator \cite{Zhang:24}. Neither of these studies utilized structured spectral illumination but this is a straightforward extension. Employing digital signal processing techniques, we show it is possible to reconstruct continuous spectra (hyperspectral) results, allowing us to retain the low data storage and throughput of multispectral imaging, but achieve hyperspectral imaging results, especially relevant for precision agriculture.

\section{Structured LED Illumination}

The reconstruction of a continuous spectrum from a finite set of overlapping broadband emitters should not be intuitively obvious. One might expect Gaussian blurring or insufficient capture of detail.  Rather than illuminating each LED one by one and recording its reflectance profile, we study the use of random structured spectral illumination, which consists of randomly turning on sets of LEDs with random intensities. The simulations used here are to find the number of LEDs needed, for a given amount of noise, to reconstruct a spectrum at a given precision using the Moore--Penrose pseudoinverse \cite{barata2012moore}, which is a least-squares solution of the reconstruction.  Our work further demonstrates that incoherent measurements can be used to reconstruct the often sparse spectra that exist in nature. This approach is computationally significantly cheaper than often-used spectral fitting approaches based on using the LED spectra as basis functions while keeping their amplitudes as free parameters (see Ref. \cite{llenas2019arbitrary}, for example). 

\begin{figure}[ht]
\includegraphics[width=.4\textwidth]{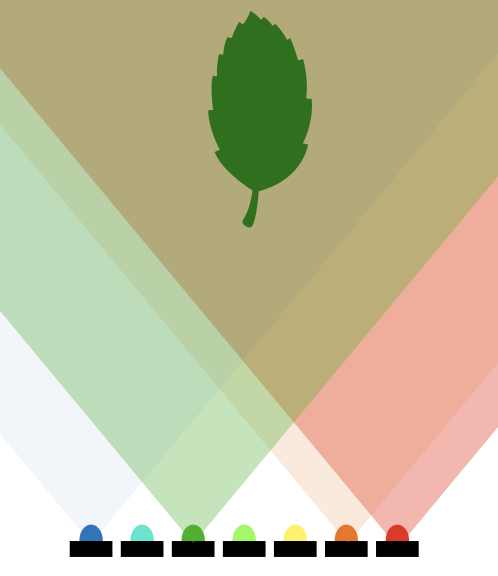}
    \caption{Schematic of the proposed experimental system.  An array of LEDs is used to illuminate an object.  For each illumination pattern, some of the LEDs are active with randomly chosen intensity.  Many different configurations of LEDs and intensities are chosen.} 
    \label{Leaf}
\end{figure}

Consider a schematic of the proposed illumination in Fig. \ref{Leaf}. An array of LEDs illuminate an object.  Some predefined fraction of the LEDs are illuminated with randomly chosen intensity. Each pixel of a camera acts as a bucket detector and records the reflectance for the particular illumination.  Each pattern is considered to be one measurement.  The process repeats with a different illumination pattern until $M$ such measurements have been taken.    

The series of random illumination patterns can be represented by a matrix $\mathbf{\Phi} \in {R}^{m\times n}$, where each of the $m$ rows of the matrix represents one spectral illumination pattern and $n$ is the length of the spectrum we wish to recreate, which can be much larger than $m$.  The random choice of LEDs and intensities allows us to create an incoherent measurement matrix and thus to use compressive sensing reconstruction models \citep{candes2008introduction, donoho2006compressed, baraniuk2007compressive,eldars2012compressed, duarte2008single}.

To analyze the information capacity of our proposed measurement matrix $\mathbf{\Phi}$, we perform Singular Value Decomposition (SVD):
\begin{equation}
    \mathbf{\Phi} = \mathbf{U} \mathbf{\Sigma} \mathbf{V}^T,
\end{equation}
which gives us the low rank approximation of the matrix.  Here, $\mathbf{U} \in {R}^{m \times m}$ and $\mathbf{V} \in {R}^{n \times n}$ are orthonormal matrices whose columns contain the left and right singular vectors, respectively. The matrix $\mathbf{\Sigma} \in {R}^{m \times n}$ is diagonal and contains the singular values $\sigma_i$ in descending order.

The rank of the matrix $\mathbf{\Phi}$ is determined by the number of nonzero singular values.  The rank specifies the number of independent spectral components captured by the illumination patterns.  However, both the rank of the matrix and the rate at which the singular values decay are important when considering the information contained within the illumination patterns.   

%The number of nonzero singular values determines the rank of $\mathbf{\Phi}$, which indicates how many independent spectral components are captured. A higher rank implies that the structured illumination patterns provide a richer spectral basis, improving the reconstruction accuracy \citep{golub2013matrix, strang1993introduction}.

Of particular importance is the role of noise in comparison to the singular values.  A standard procedure to deal with noise is to rewrite the singular value decomposition as a sum of matrices weighted by singular values,
\begin{equation}
    \mathbf{\Phi} = \sum_{i=1}^{r} \sigma_i \mathbf{u}_i \mathbf{v}_i^T, \label{truncatedSeries}
\end{equation}
where $\mathbf{u}_i$ are the left singular vectors or the columns of $\mathbf{U}$ and $\mathbf{v}_i$ are the right singular vectors or columns of $\mathbf{V}$.  The matrix for each singular value is found by taking the outer product of $\mathbf{u}_i$ and $\mathbf{v}_i^T$.   One can truncate this sum and retain only those terms where $\sigma_i$ is greater than some threshold $\tau$ set by the noise. 

We represent the measurement vector $\mathbf{y} \in {R}^m$ by
\begin{equation}
    \mathbf{y} = \mathbf{\Phi} \mathbf{x} + \mathbf{w},
\end{equation}
where $\mathbf{x} \in {R}^{n}$ is the reflectance spectrum to be found, and $\mathbf{w} \in {R}^{m}$ is modeled by additive Gaussian white noise. The measurement vector $\mathbf{y}$ is the total integrated flux measured in each pixel of a camera.   We seek to find $\mathbf{x}$, or at least a good approximation, from $\mathbf{y}$. 

Our interest is in reconstructing spectra having sparse realizations in an orthonormal basis. Stated differently, if the amount of information in the structured illumination is close to or exceeds the information content of the spectrum, it is possible to obtain a high quality approximation of the spectrum.      

\section{Spectrum Reconstruction}
We find \(\mathbf{x}\) from \(\mathbf{y}\), using the inverse of the singular value decomposition. When $\Phi$ is rank deficient or rectangular (underdetermined), we take the Moore--Penrose pseudoinverse, which is given by the inverse of the singular value decomposition used extensively in many branches of science \cite{klema1980singular,stewart1993early,miller2000communicating,wall2003singular}.  For underdetermined systems, the Moore--Penrose pseudoinverse finds the solution with the smallest Euclidean norm :
\begin{equation}
    \min_{\mathbf{x}} \|\mathbf{x}\|_2 \quad \text{s.t.} \quad \textbf{y}=\Phi\textbf{x}
\end{equation}
which has a solution given by
\begin{equation}
    \mathbf{x'}=\mathbf{\Phi}^+\mathbf{y},
\end{equation}
where
\begin{equation}
    \mathbf{\Phi}^{+} = \mathbf{V} \mathbf{\Sigma}^{+} \mathbf{U}^T. 
\end{equation}
In this pseudoinverse, \(\mathbf{\Sigma}^{+}\) is obtained by inverting the nonzero singular values.  For reasons of noise mitigation, we use the truncated form of the matrix $\mathbf{\Phi}$ in Eqn. \ref{truncatedSeries} to find the inverse matrix, namely
\begin{equation}
    \mathbf{\Phi}^{+} = \sum_{i=1}^{r} \frac{1}{\sigma_i} \mathbf{v}_i\mathbf{u}_i^T . \label{truncatedInverse}
\end{equation}
Since the singular values are inverted ( i.e., $\frac{1}{\sigma_i}$), small singular values can amplify noise. For this work, we truncate the series when the standard deviation of the noise is larger than the singular value. \citep{hansen1987truncated, hansen1990truncated}.

\section{Simulations}
To simulate the behavior of the system, we make a few simplifying assumptions.  We assume that the spacing between and the bandwidth of the LEDs are the same for each LED.  In reality, the bandwidth tends to grow with increasing wavelength in the visible and near-infrared.  Further, while there are hundreds of different LEDs with different center wavelengths spanning this range, the spectral separation is not uniform.  We also assume that the detector has a uniform response from 400 nm to 1000 nm.  This is obviously not true, but the behavior can be compensated by knowing the spectral response of a given detector.  

\begin{figure}[h]
\includegraphics[width=.47\textwidth]{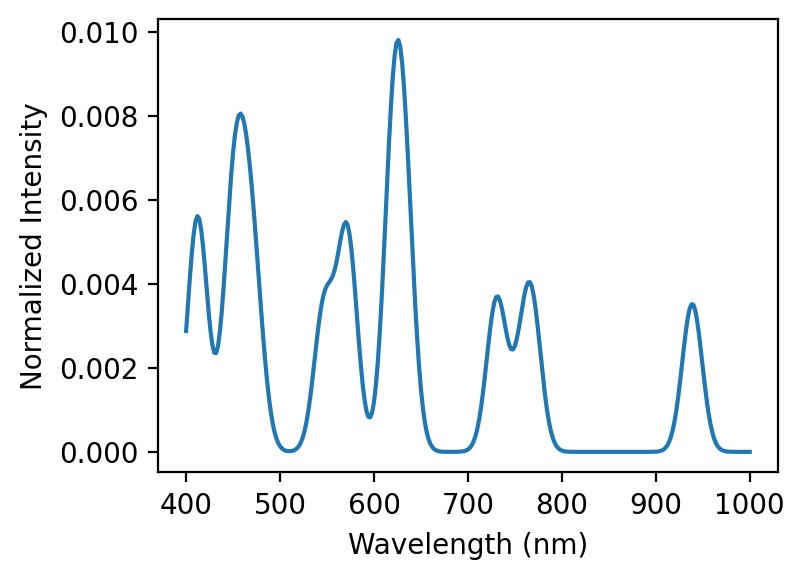}
    \caption{An example of the spectrum of one illumination pattern.  In this example, 50 LEDs, each of 15 nm bandwidth, are in an array, but only 10 of them are active, each having a randomly chosen intensity.  The normalization was performed by integrating over the 1D pattern and then dividing all terms by the sum.} 
    \label{Pattern}
\end{figure}

Consider an example of a single spectral illumination pattern shown in Fig. \ref{Pattern}. This example assumes that there are 50 LEDs used in the experiment, each with a Gaussian bandwidth of $\sigma_i=$15 nm  and spectral bins of 2 nm width.  One-fifth of the LEDs are illuminated, each LED having a random intensity.  The random illumination pattern, shown in Fig. \ref{Pattern}, is normalized by dividing by the sum of the intensities in all spectral bins. It is clear in the image that the spacing between the LEDs is smaller than the bandwidth. This example corresponds to a single row in the measurement matrix $\mathbf{\Phi}$. In this paper, we explore both the bandwidth and the number of LEDs to reconstruct the spectra for a given amount of noise.  

Next, we quantize the information capacity of a measurement matrix represented by many such illumination patterns.  If $q$ is the number of LEDs, the number of patterns must be greater than the number of random measurement patterns, i.e., $m\geq q$. Otherwise, the rank of the matrix is set by the number of patterns and not by the number of LEDs.  While compressive reconstruction affords reduced construction time ($m\ll n$), we do not consider it to be critical when using the pseudoinverse.  Thus, for this work, we used $n\geq m\geq q$. 

\begin{figure}[ht]
\includegraphics[width=.47\textwidth]{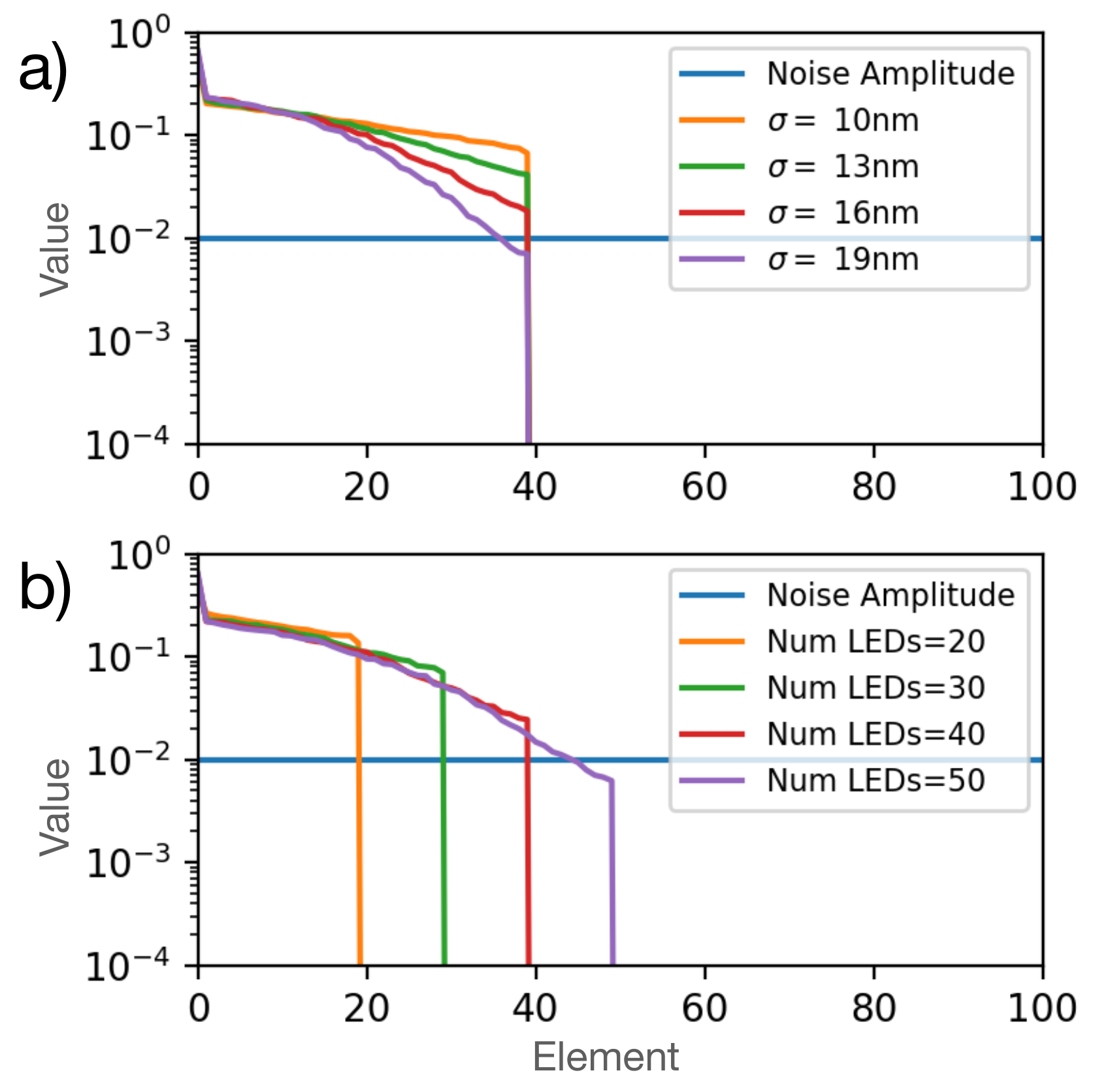}
    \caption{ A graphical form of the  singular value matrix $\mathbf{\Sigma}$ is shown. The element of the matrix is shown on the x-axis with its normalized value on the y-axis.  In a) the singular values of $\mathbf{\Sigma}$ are shown for changing LED bandwidth.  It clearly shows that the singular values drop precipitously with increasing bandwidth. The measurement matrix is normalized to the Frobenius norm and the example Gaussian white noise is scaled to $1\%$ (horizontal blue line).  In b) the singular values for $\sigma=15$ nm are shown for 20, 30, 40, or 50 uniformly spectrally spaced LEDs. It is clear that the rank of the matrix is equal to the number of LEDs. The measurement matrix is normalized to the Frobenius norm and the example Gaussian white noise is scaled to $1\%$ (horizontal blue line). }   
    \label{SvLED}
\end{figure}

Consider the measurement matrix quantization results shown in Fig. \ref{SvLED} b).  Here, the singular values of the submatrix $\Sigma$ are plotted for 20, 30, 40, or 50 uniformly displaced LEDs with 15 nm bandwidth.  The singular values are properly normalized using the Frobenius norm on the matrix $\mathbf{\Phi}$ \cite{stewart1993early}.  We chose a Gaussian white noise threshold to be $1\%$ on this scale (shown in blue). It is clear from the figure that the number of non-zero singular values, i.e., the rank of the measurement matrix, is equal to the number of LEDs.     

The effect of changing the bandwidth on the singular values is shown in Fig. \ref{SvLED} a).  Importantly, one can see that the singular values drop much more rapidly with increasing LED bandwidth, making noise much more important.  This is an intuitive result.  For 50 LEDs, the average spacing is about 12 nm.  A bandwidth of $\sigma=19$ nm means that the separation between the nearest neighbors is much smaller than the bandwidth.  One would expect a rapidly diminishing response.     

\section{Reconstruction examples}

\begin{figure}[ht]
\includegraphics[width=.47\textwidth]{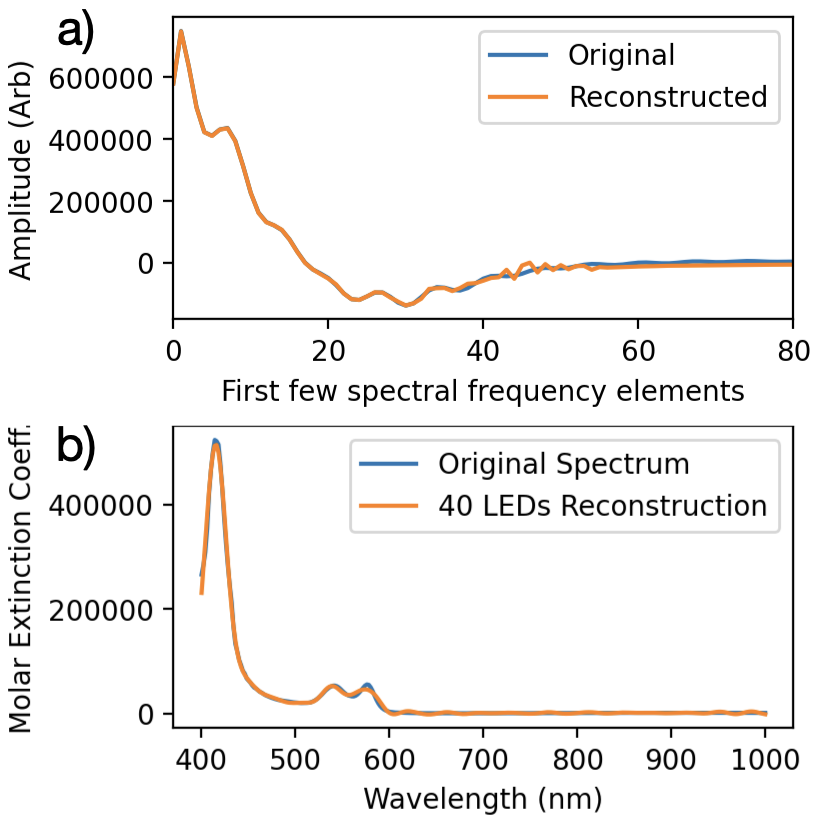}
    \caption{Oxyhemoglobin spectra vs example reconstruction. a) shows the first few elements of the discrete cosine transform of the spectra and the reconstruction of the DCT using the Moore-Penrose inverse.  It can be seen that the DCT of the original spectrum and the reconstructed DCT remain close until the size of the singular values approaches the magnitude of the noise.   b) shows the reconstruction vs the original oxyhemoglobin spectrum. } 
    \label{Hemoglobin}
\end{figure}

\begin{figure}[ht]
\includegraphics[width=.47\textwidth]{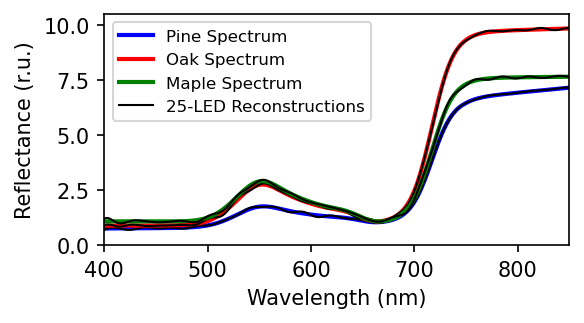}
    \caption{Reflectance spectrum of pine branches (blue), oak leaves (red), and maple leaves (green) vs each spectrum's reconstruction (black). The reconstructed spectra used 25 LEDs and a noise of 1\%.} 
    \label{COMSPECT}
\end{figure}

As representative spectral curves, we study the reconstruction of the oxyhemoglobin spectrum (Fig. \ref{Hemoglobin}) and three empirical reflectance spectra of green vegetation (Fig. \ref{COMSPECT}). As can be seen, both types of spectra are inherently sparse.     

In the oxyhemoglobin example, we use 40 LEDs, each having a bandwidth of 15 nm and a measurement noise of 1\%.  From Fig. \ref{SvLED}, we can see that when the size of the singular values is of the order of the noise, we should expect poor reconstruction.  In Fig. \ref{Hemoglobin}, we decomposed the oxyhemoglobin spectrum using the discrete cosine transform (DCT) as well as the reconstructed spectrum using the Moore-Penrose pseudoinverse.  Around element 40, we begin to see appreciable noise, as expected from the study of the singular values.  

\begin{figure}[ht]
\includegraphics[width=.47\textwidth]{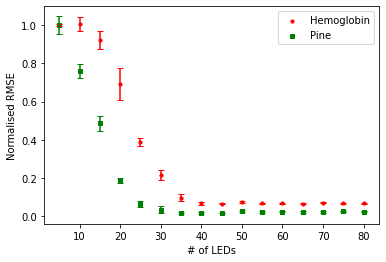}
    \caption{The normalized RMSE values between the reconstructed and original oxyhemoglobin and pine spectra were calculated  for a varying number of LEDs. Each reconstruction was performed 10 times with random illumination patterns. RMSE values were normalized to allow for direct comparison.} 
    \label{RMSE_LEDs}
\end{figure}

The fidelity of the oxyhemoglobin and pine reflectance reconstruction as a function of the number of LEDs was tested and the results depicted in Fig. \ref{RMSE_LEDs}. From these results, it can be clearly seen that for oxyhemoglobin 40 LEDs are sufficient to near perfectly reconstruct the reflectance spectrum, while 25 LEDs are sufficient to reconstruct the pine spectrum with high fidelity.

The green tree spectra were simulated using the empirical model COMSPECT and fitting parameters given in Ref. \citep{COMSPECT}. These spectra were reconstructed well using only $\sim$25 LEDs. The contributions from individual Gaussians are visible on the red edge of the spectra. This is expected since the spectral amplitudes increase until the spectra are cropped at 900 nm.  This shows that such a system can be used in precision agriculture \cite{behmann2014detection,rajkumar2012studies}.  

\section{Discussion}
We have focused on the simulations of randomly illuminated LED spectra.  Performing the experiment should be straightforward, but with attention to detail. There are five primary experimental concerns.  First, we have assumed that the response spectrum of the detector is flat.  To compensate for nonideal behavior in a standard camera, the responsivity at each wavelength must be determined and the LED illumination intensity at each wavelength must be increased or decreased accordingly. This will create the effect of a flat spectral response. Second, the actual spectrum of each LED must be measured and incorporated in the measurement matrix.  Third, to adjust the relative intensity pattern, one can incorporate a digitally tunable resistor network or use pulse-width modulation, which is fast compared to the exposure time of the camera. Fourth, the LEDs will need to have illumination patterns that are fairly uniform across the imaging plane, so that illumination from different source locations will not affect the results.  We have found that surface-mounted LED designs have very uniform illumination. Fifth, it can be difficult to find LEDs of the same type in some spectral regions, which might require using different kinds of LEDs, power supplies, etc.  This will change the assumption about uniform spacing of the LEDs, but can be corrected with the knowledge of the individual LED spectra.  We note that DigiKey has many dozens of colors from the ultraviolet to well into the infrared.  However, we believe all of these problems are tractable and in line with our previous experimental work \cite{howell2024raspberrypimultispectralimaging}.            

The purpose of this work was to demonstrate that randomly illuminated LED spectral patterns could be used to reconstruct continuous spectra using signal processing methods.  Owing to the low cost and broad range of available LED wavelengths in the visible and infrared, such a study was warranted.  The use of the Moore-Penrose inverse was to tease out the information density in such spectral illumination patterns, but not necessarily as a reconstruction method. Fortunately, in this case, the inverse was sufficient to reconstruct.  However, more importantly, it shows that better optimized signal reconstruction methods in compressive sensing and digital signal processing can take advantage of these random illumination properties to efficiently match the continuous spectrum. 

While not the focus of this paper, we note that we also performed spectral reconstruction simulations using Orthogonal Matching Pursuit, following the approach of \cite{mallat1993matching,tropp2007signal}, to obtain similar reconstruction results, but with far fewer random illumination patterns. We found that we only needed the same number of patterns as there were singular values (i.e., $m=q$ as described in section IV). 

The reconstruction of several green vegetation spectra with as few as 25 LEDs is an encouraging finding. It implies that such a camera can be used to great success within the precision agriculture domain.  Importantly, unlike a standard camera, which is intended to be general-purpose in that it can take a picture of any possible scene, these multi-spectral cameras tend to be purpose-built.  Hence, in a focused study, it is usually more important to look at changes in one spectrum (i.e., difference detection) than it is to look at a broad set of spectra.  Thus, we expect a multi-spectral camera that can be used to reconstruct continuous spectra with relatively few LEDs can play an important role in focused research domains.  

Lastly, the continuous reconstruction from a few finite measurements is only made possible by the sparsity of the signal in some basis.  We used the discrete cosine transform as a representative basis.  As long as the number of sparse elements in the spectrum is less than or equal to the number of singular values, a good reconstruction is possible.  Therefore, an important problem is not only increasing the number of singular values, but also finding the basis in which a spectrum is the sparsest, allowing for optimized reconstruction. 

\section{Conclusions}
In conclusion, we have studied the use of digital signal processing for reconstructing continuous spectra using structured illumination of LEDs.  Owing to the sparsity of most spectra, especially in the agricultural domain, we believe that a low-cost camera composed of LEDs or finite filter sets can perform the imaging tasks of much more expensive hyperspectral cameras.

\section{Acknowledgments}
JCH acknowledges support from Chapman University.

\vfill{\eject}

\bibliography{Main}
% \clearpage
% \appendix
% \onecolumngrid
% \input{supplemental.tex}
\end{document}